\documentclass[onecolumn,pre]{revtex4}
\usepackage{bm}
\usepackage{epsfig}
\usepackage{amsmath}
\begin{document}
\title{Barrierless Reactions in Solution: An Analytically Solvable Model}
\author{Aniruddha Chakraborty \\
School of Basic Sciences, Indian Institute of Technology Mandi,\\
Mandi, Himachal Pradesh, 175001, India}
\date{\today }
\begin{abstract}
\noindent  We propose an analytical method for solving the problem of barrierless reactions in solution, modeled by a particle undergoing diffusive motion under the influence of both reactant and product potentials. The coupling between these two potentials is taken to be a Dirac Delta function. The diffusive motion in this paper is described by the Smoluchowskii equation. Our solution requires only the knowledge of the Laplace transform of the Green's function for the motion in both the uncoupled potentials. Our model is more general than all the earlier models, because we are the first one to consider the potential energy surfaces of both the reactant and product explicitly.
\end{abstract}
\maketitle
Many important chemical reactions do not face any activation barrier in their reactive motion along the reaction coordinate from the reactant to product \cite{BagchiBook,Marcus,Agmon}. The dynamics of those barrierless reactions differ considerably from chemical reactions where reactants has to overcome an activation barrier \cite{Shank,Hoffmann,Cremers,Harris,Nishijima,Fleming,Oxtoby}. In general barrierless reactions are very fast. An important example is vision transduction process, which involves a barrierless cis to trans transformation of  rhodopsin \cite{Shank}. Other important examples include isomerization of stilbene and diphenyl butadiene in solution \cite{BagchiBook}. The experimental observable in most of the reactions is the reactant survival probability, which is time dependent. Time dependence of reactant in solution is often viscosity dependent \cite{Shank,Hoffmann,Cremers,Harris,Nishijima,Fleming,Oxtoby}. This usually indicates that the reactive motion is a diffusive motion along the reaction coordinate \cite{BagchiBook}. There have been many theoretical research to explain the experimental observations \cite{BagchiBook,Nishijima,Fleming,Oxtoby}. In the absence of an activation barrier, there is no clear separation of time scales between the motion in the reactive region and in the other part of potential energy surface. So steady state solution is no longer possible and one has to solve the full time dependent probability distribution of the reactant on the reactant potential energy surface and the same for product on product potential energy surface \cite{BagchiBook}. The typical potential energy surface for barrierless reaction considered in this paper is the following. There is a sizable activation barrier between the reactant and product, in the ground state potential energy surface. So reactant will face this activation barrier in their reactive motion along the reaction coordinate. But there is no activation barrier between the reactant and product, in the excited state potential energy surface. So the reactant is excited from the ground state to the excited state potential energy surface by light. The subsequent relaxation brings the reactant down towards the product region of the potential energy surface. In this region, if the system is de-excited to the ground state potential energy surface, we have the desired product. The experimental observable in most of the reactions is the system's survival probability in excited potential energy surface, which is denoted here as $P_{e}(t)$, the more fundamental quantity here is $P_e(x,t)$, which provides the position dependent population distribution on the excited state potentiale energy surface, where $x$ denotes the location of the particle along the reaction coordinate. The survival probability $P_e(t)$ is defined using $P_e(x,t)$ by averaging over the entire reaction coordinate $x$.
\begin{equation}
P_e(t)=\int dx P_{e}(x,t).
\end{equation}
Theories of barrierless reactions are based on a simple description of the motion of the system on the excited potential energy surface and the reaction is described by a position dependent sink term along the reaction coordinate. This can be described by the following equation.
\begin{equation}
\frac{\partial P_e(x,t)}{\partial t} = {\cal L}_e P_e(x,t) - k_r P_e(x,t) - k_0 S(x) P_e(x,t),
\end{equation}
where $x$ is the reaction coordinate and $\cal L$ is the Smoluchowskii operator. The third term in the right hand side is usually referred to as sink term, which is the reason for the decay of reactant population at $x$. In most of the cases the position dependence of $S(x)$ makes the above equation quite difficult to solve analytically. The problem essentially is to calculate the probability that the system will still remain in the excited potential energy surface after a finite time $t$. Many simple models have been proposed to solve this equation analytically or numerically  \cite{BagchiBook,Nishijima,Fleming,Oxtoby}. It is interesting to note that none of the theoretical studies so far, have considered any effect of ground potential energy surface on $P_{e}(x,t)$. We denote  the probability that the system would survive on the excited potential energy surface by $P_{e}(x,t)$. We also use $P_{g}(x,t)$ to denote the probability that the system would be found in the ground  potential energy surface. Both the probability $P_{e}(x,t)$ and $P_{g}(x,t)$ may be found at $x$ at the time $t$ obeys a modified Smoluchowskii equation.
\begin{eqnarray}
\frac{\partial P_e(x,t)}{\partial t} = {\cal L}_e P_e(x,t) - k_r P_e(x,t) - k_0 S(x) P_g(x,t)  \\ \nonumber
\frac{\partial P_g(x,t)}{\partial t} ={\cal L}_g P_g(x,t) - k_r P_g(x,t) - k_0 S(x) P_e(x,t). \nonumber
\end{eqnarray}
In the above 
\begin{equation}
{\cal L}_i= A \frac{\partial^2}{\partial x^2}+\frac{\partial}{\partial x} \frac{dV_i(x)}{dx}.
\end{equation}
$V_i(x)$ is the potential causing the drift of the particle, $S(x)$ is a position dependent sink function, $k_0$ is the rate of nonradiative decay and $k_r$ is the rate of radiative decay. We have taken $k_r$ to be independent of position. $A$ is the diffusion coefficient. Our model is more general than all the earlier models  \cite{BagchiBook,Nishijima,Fleming,Oxtoby}, in the sense that our model also considers the motion on the ground state potential energy surface. It is quite unlikely that the shape of the ground state potential energy surface do not play any role, but unfortunately this fact was not considered in any of the earlier studies \cite{BagchiBook,Nishijima,Fleming,Oxtoby}.
Before we excite, the system is in the ground state potential energy surface, and as the solvent is at a finite temperature, its distribution over the coordinate $x$ is random. From this it undergoes Franck-Condon excitation to the excited state potential energy surface. So, $x_0$ the initial position of the particle, on the excited state potential energy surface is random. We assume it to be given by the probability density $P^{0}_{e} (x_0)$. In the follwoing we provide a general procedure for finding the exact analytical solution of Eq. (1). The Laplace transform ${\cal P}_i(x,s)=\int_{0}^{\infty} P_i(x,t)e^{-st} dt$ obeys
\begin{eqnarray}
[s-{\cal L}_e+k_r] {\cal P}_e(x,s)+k_0 S(x) {\cal P}_g (x,s) = P^0_e(x_0) \\ \nonumber
[s-{\cal L}_g+k_r] {\cal P}_g(x,s)+k_0 S(x) {\cal P}_e (x,s) = 0, \nonumber
\end{eqnarray}
where $P^0_e(x_0)=P_e(x,0)$ is the initial distribution at the excited state potential energy surface and $P_g(x,0)=0$ is the initial distribution at the ground state potential energy surface. 
\begin{equation}
 \left(
\begin{array}{c}
{\cal P}_e (x,s) \\
{\cal P}_g (x,s)
\end{array}
\right) = \left(
\begin{array}{cc}
s-{\cal L}_e+k_r & k_0 S(x) \\
k_0 S(x) & s-{\cal L}_g+k_r
\end{array}
\right)^{-1}
\left(
\begin{array}{c}
P^0_e(x) \\
0
\end{array}
\right)  ,
\end{equation}
Using the partition technique \cite{Lowdin}, solution of this equation can be expressed as 
\begin{equation}
{\cal P}_e(x,s)=\int_{-\infty}^{\infty} dx_0 G(x,s;x_0)P^0_e(x_0),
\end{equation}
where $G(x,s;x_0)$ is the Green's function defined by
\begin{equation}
G(x,s;x_0)=\left < x \left|[s-{\cal L}_e+ k_r - {k_0}^2 S[s-{\cal L}_g+k_r]^{-1}S]^{-1}\right| x_0 \right>
\end{equation}
The above equation is true for any general $S(x)$. This expressions simplify a lot if $S(x)$ is a Dirac Delta function located at $x_c$. In operator notation $S$ may be written as $S= \left| x_c \left> \right < x_c \right |$. Then
\begin{equation}
G(x,s;x_0)=\left < x \left|[s-{\cal L}_e+ k_r - {k_0}^2 G^{0}_g(x_c,s;x_c) S ]^{-1}\right| x_0 \right>,
\end{equation}
where
\begin{equation}
G^{0}_g(x,s;x_0)=\left < x \left|[s-{\cal L}_g+ k_r ]^{-1}\right| x_0 \right>
\end{equation}
and corresponds to propagation of the particle starting from $x_0$ on the ground state potential energy surface in the absence of coupling.
Now we use the operator identity
\begin{equation}
[s-{\cal L}_e + k_r - {k_0}^2 G^{0}_g(x_c,s;x_c) S]^{-1}=[s-{\cal L}_e+ k_r]^{-1}-[s-{\cal L}_e+ K_r]^{-1}{k_0}^2 G^{0}_g(x_c,s;x_c) S [s-{\cal L}_g + k_r - {k_0}^2 G^{0}_g(x_c,s;x_c) S]^{-1}
\end{equation}
Inserting the resolution of identity $I=\int_{-\infty}^{\infty} dy \left|y \left> \right < y \right|$ in the second term of the above equation, we arrive at an equation which is similar to Lippman-Schwinger equation.
\begin{equation}
G(x,s;x_0)=G^e_0(x,s;x_0) + {k_0}^2 G^0_e(x,s;x_c)G^0_g(x_c,s;x_c)G(x_c,s;x_0).
\end{equation}
where $G^0_e(x,s;x_0)=\left < x \left|[s-{\cal L}_e+k_r]^{-1}\right| x_0 \right>$ corresponds to the propagation of the particle put initially at $x_0$, in the absence of coupling, it is actually the Laplace Transform of $G_0(x,t;x_0)$, which is the probability that a particle starting at $x_0$ can be found at $x$ at time $t$. We now put $x=x_c$ in the above equation and solve for $G(x_c,s;x_0)$ to get
\begin{equation}
G(x,s;x_0)=G^e_0(x,s;x_0)\left(1- {k_0}^2 G^0_e(x_c,s;x_c)G^0_g(x_c,s;x_c)\right).
\end{equation}
This when substitued back into Eq. (10) gives
\begin{equation}
G(x,s;x_0)=G^e_0(x,s;x_0) + \frac{{k_0}^2 G^0_e(x,s;x_c)G^0_g(x_c,s;x_c)G^0_e(x_c,s;x_0)}{1-{k_0}^2 G^0_e(x_c,s;x_c)G^0_g(x_c,s;x_c)}.
\end{equation}
Using this Green's function in Eq. (5) one can caluclate ${\cal P}_e(x,s)$ explicitely. Here we are interested to know the survival probability at the excited state potential energy surface $ P_e(t) = \int_\infty^\infty dx P(x,t)$. It is possible to evaluate Laplace Transform  ${\cal P}_e(s)$ of $P_e(t)$ directly. ${\cal P}_e (s)$ is defined in terms of ${\cal P}(x,s)$ by the following equation,
\begin{equation}
{\cal P}_e(s)=\left(1-\left[1+k_0^2 G^0_e(x,s;x_c)G^0_g(x_c,s;x_c)\right]^{-1}k_0^2 G^0_g(x_c,s;x_c)\int^{\infty}_{-\infty}dx_0 G^0_e(x_c,s;x_0)P^0_e(x_0)\right)/(s+k_r).
\end{equation}
From the above equation we see that ${\cal P}_e(s)$ depends on $G^0_g(x_c,s;x_c)$ which is different from the results of all earlier studies   \cite{BagchiBook,Nishijima,Fleming,Oxtoby}. The average and long time rate constants can be found from ${\cal P}_e(s)$. Thus, $k^{-1}_{1}={\cal P}_e(0)$ and $k_{L}= - ($ pole of $\left[1+k_0^2 G^0_e(x,s;x_c)G^0_g(x_c,s;x_c)(s+k_r)\right]^{-1})$, closest to the origin, on the negative $s$ - axis, and is independent of the initial distribution but depends on $G^0_g(x_c,s;x_c)$. The expression that we have obtained for ${\cal P}_e(s)$, $k_I$ and $k_L$ are quite general and are valid for any set of potentials. \par
Barrierless chemical reaction are actually very common in nature \cite{BagchiBook}. These reactions proceed at an astonishingly fast rate. Interesting results are obtained from experiments for this type of chemical reactions, such as nonexponential decay kinetics. There are standard models available to expalin the experimental results, essentially from nonsteady-state populations dynamics on the reactant potential energy surface, but the solution of those relevant equations were done mostly by numerical methods. In this paper we have generalized standard model with inclusion of effects from non-steady state population dynamics on the product potential energy surface too. We showed that this model can even be solved exactly by analytical method.

\end{document}